\documentclass{article}
\usepackage{natbib, epsfig, emulateapj, apjfonts}
\def\be{\begin{equation}}
\def\ee{\end{equation}}

\newbox\grsign \setbox\grsign=\hbox{$>$} \newdimen\grdimen \grdimen=\ht\grsign
\newbox\simlessbox \newbox\simgreatbox \newbox\simpropbox
\setbox\simgreatbox=\hbox{\raise.5ex\hbox{$>$}\llap
     {\lower.5ex\hbox{$\sim$}}}\ht1=\grdimen\dp1=0pt
\setbox\simlessbox=\hbox{\raise.5ex\hbox{$<$}\llap
     {\lower.5ex\hbox{$\sim$}}}\ht2=\grdimen\dp2=0pt
\setbox\simpropbox=\hbox{\raise.5ex\hbox{$\propto$}\llap
     {\lower.5ex\hbox{$\sim$}}}\ht2=\grdimen\dp2=0pt

\begin{document}

\title{
Intermediate-mass Black Hole(s) and stellar orbits in the
Galactic Center
}

\author{ Yuri Levin, Alice Wu, and Ed Thommes}

\affil{Canadian Institute for Theoretical Astrophysics,
60 St. George Street, Toronto, ON M5S 3H8, Canada}

\begin{abstract}
Many young stars reside within the central half-parsec from
SgrA*, the supermassive black hole in the 
Galactic Center. The origin of these stars remains
a puzzle.
Recently, Hansen and Milosavljevic (2003, HM) have argued
that   an 
Intermediate-Mass Black Hole (IMBH) could have  delivered  the young stars to
the immediate vicinity of SgrA*.  Here we focus on the final stages of the
HM scenario. Namely, we integrate numerically the orbits of stars which are initially
bound to the IMBH, but are  stripped from it by the tidal field of SgrA*. 
Our numerical algorithm is a symplectic integrator designed  specifically for
the problem at hand; however, we have checked our results with 
SYMBA, a version of the widely available SWIFT  code.  We find that  the 
distribution of the post-inspiral orbital parameters
is sensitive to the eccentricity of the inspiraling IMBH.
If the IMBH is on a circular orbit, then
the inclinations of numerically computed orbits relative to the inspiral plane 
 are almost always smaller than 10 degrees, and 
therefore (a) the simulations are in good agreement with the observed motions
of stars in a clockwise-moving
stellar disc, (b) the simulations never reproduce the orbits of stars outside this disc, which
include those in the second thick ring of stars and the randomly oriented unrelaxed orbits of some of
the S-stars. If the IMBH's orbital eccentricity is $e=0.6$, then approximately
half of the stars end up with orbital inclinations below 10 degrees, and another
half have inclinations anywhere between $0$ and $180$ degrees; this is somewhat
closer to what's observed. We also show
 that if IRS13 cluster is bound by an IMBH, as has been argued by Maillard et.~al.~2004,
then the same IMBH could
not have delivered all of the young stars to their present location. 

\end{abstract}



\section{Introduction}
A few tens of bright, young, and massive stars have been observed in
close proximity ($0.001$---$0.5$ pc) to the radio source SgrA*, which is associated with
the supermassive ($4\times10^6M_{\odot}$) black hole in the Galactic Center\footnote{For convenience,
we  shall
refer to the supermassive black hole simply as SgrA*.} (see Genzel et.~al.~2000,
Ghez et.~al.~,2003, Shodel et.~al.~2002,2003 for recent work). 
The apparent youth of these  stars implies two logical possibilities
for  their birth history: (1) either they were born {\it in situ}, in the immediate vicinity
of SgrA*, or (2)  they were born some distance away and then  delivered to SgrA* within a few million years after their
birth. In the first scenario, 
the {\it in situ} birth would occur in a strong tidal field of SgrA*, and the density of the star-forming gas
would have to be several orders of magnitude greater than that anywhere else in the Galaxy (Phinney 1989,
Sanders 1992, Morris 1993). The most likely geometry for the star-forming gas is a dense disc, which is 
either compressed by the central explosion powered by SgrA* (Morris, Ghez, and Becklin 1999), or, more plausibly, which is massive enough
to become self-gravitating and fragment into stars\footnote{We note that it is by no means certain
that the disc would circularize before forming stars (J. Goodman, private communications).}
 (Levin and Beloborodov 2003, Nayakshin, Cuadra, and Sunyaev 2004,
Milosavljevic and Loeb 2004, Nayakshin and Cuadra 2004).  
The {\it in situ} formation is not the focus of this paper, although we will 
briefly return to it in  our concluding remarks.

The  second scenario, which is the focus of this paper, calls
 for rapid delivery of the young stars to the supermassive Black Hole. 
The initial version of it was proposed by Gerhard 2001, and investigated in
more detail by McMillan and Portegies Zwart 2003, and Kim and Morris 2002.
 Gerhard's basic idea was that if a massive ($\sim 10^6M_{\odot}$) cluster
of stars was born $10$---$30$pc away from SgrA*, then the dynamical friction in the central bulge would
bring the cluster towards  SgrA* on the timescale of a few million years. During the inward migration, the massive stars in the cluster
would sink to the cluster's center and form a compact core. Gerhard has argued that even though the envelope of the
cluster gets stripped by the tidal field a few parsecs away from SgrA*, the dense cluster core would only be tidally disrupted
within the central parsec. However, even that is not close enough: the observed stars reside $0.001$---$0.5$ pc from SgrA*
and the core density would have to increase by orders of magnitude before it could  get so close to SgrA*. 

Recently, Hansen and Milosavljevic 2003 (from now on, HM) have suggested a fix to Gerhard's scenario which would enable the
stars to be delivered to the observed proximity from SgrA*. They have argued that if an Intermediate-Mass Black Hole (IMBH), with
the mass of $10^3$---$10^4M_{\odot}$, was positioned at the center of the cluster, than it would provide some extra gravitational
binding for the cluster core. This scenario is supported by an observation
of the seemingly bound mini-cluster IRS13 at $\sim 0.1$pc from SgrA* (Maillard et.~al.~2004,
from here on MPSR).  MPSR have argued that the presence of an IMBH at the
center of IRS13 is the most plausible explanation of how the mini-cluster
keeps itself together in the strong tidal field of SgrA*.  On the theoretical side,
a number of investigations 
argue that IMBH would plausibly form 
in a dense core of a young very massive cluster
(Portegies Zwart et.~al.~2004, Gurkan et.~al.~2004).
 The  cluster+IMBH inspiral would  consist of two stages: the first one, in which most of the most of
the cluster is tidally disrupted at about a parsec from
SgrA* and the IMBH is released together with a few tens of massive stars tightly bound to it, and the
second one, in which the IMBH continues its inspiral   
 while the stars are peeled off one-by-one by the SgrA* tidal pull, and are eventually put into
nearly Keplerial orbits around SgrA*.  The first stage was recently investigated numerically by 
Kim and Morris 2003 (KM) and by Gurkan and Rasio 2004. KM 
have found that very few stars survive bound to IMBH once it gets to the central parsec. 
However, one of the numerical
limitations in KM (pointed out by the authors) was the 
large softening length for gravitational interactions between the stars.
Thus KM did not model faithfully the internal dynamics
of the  cluster. A more exact Monte-Carlo-based simulation of Gurkan and Rasio 2004 
has found that the number of stars bound to IMBH at the end of the
first stage is consistent with the number of young stars in the central half-parsec.

Here we focus on the second stage---the final inspiral of the IMBH.  We investigate 
numerically the stellar
orbits produced by such an inspiral, and compare them to observations. 
In the next section we describe the  algorithm which was used for the 
numerical orbital integrations; this
section may be skipped by a reader uninterested in technical details\footnote{In our view,
however, this algorithm is of interest in its own right and may be used in other contexts.}.  
In section 3 we present our numerical results, and in section 4 compare them to the data. We find that while 
the inspiral on a circular orbit 
explains nicely the clockwise-moving stellar disc in the 
Galactic Center (Levin and Beloborodov 2003, Genzel et.~al.~2003), it does not reproduce the
thick counterclockwise stellar disc and the randomly oriented tight orbits of the S-stars 
(Ghez et.~al.~2003,
Shodel et.~al.~2003). The inspiral on an eccentric orbit
does push some orbits out of the inspiral plane, but is still not
entirely consistent with the current data. We also show that the proposed IMBH at the
center of IRS13 almost certainly could not have delivered all of the young stars
to the Galactic Center.
Finally, we speculate how the HM scenario may be modified to account more
closely for all of the kinematic   data.

\section{Our numerical algorithm: symplectic integrator in the extended phase space}
The stars in our system will experience repeated close encounters with the IMBH even after they get tidally
stripped from it, and we expect the resulting orbits to be very chaotic and sensitive to the initial conditions.
Therefore, our study will be statistical in nature: we are seeking to find the {\it distribution} of orbits 
which are produced by the IMBH inspiral. Symplectic algorithms (SA; see Yoshida 1990) are particularly 
suitable for statistical studies,
since they generate trajectories which conserve the phase-space volume exactly. Integrators which use 
SAs  have excellent long-term
behavior, since the Hamiltonian mapping used in a SA is integrated exactly and is only slightly
different from a true Hamiltonian mapping of the system. However, for our 
purposes the simplest symplectic codes have a disadvantage since they do not handle well
the close encounters between a star and IMBH (see Preto and Tremaine 1999 for discussion).  The origin
of this limitation is the reduced timestep which is used to integrate through the
encounter; reduction of the timestep leads to change of the SA Hamiltonian and thus may cause spurious energy
changes. To overcome this problem, Mikkola (1997), 
Preto and Tremaine (1999), Mikkola and Tanikawa (1999), and Mikkola and Wiegart (2002) have developed
{\it extended phase-space} Symplectic Integrators which
use fixed timestep but in an extended phase space. The basic idea for extending the phase space of the Hamiltonian
system goes back to Poincare, and can be briefly described as follows:

Let $H(q,p,t)$ be a time-dependent Hamiltonian, where $q$ and $p$ stand for spatial coordinates and
their conjugate momenta respectively and $t$ is the time. Let us extend the phase space of the system
by another coordinate $q_0$ and its conjugate momenta $p_0$.  Consider now a new Hamiltonian
$\Gamma$ in the extended phase space (Mikkola 1997):
\begin{equation}
\Gamma(q,q_0,p,p_0)=g(q,q_0,p,p_0)\left[H(q,p,q_0)+p_0\right],
\label{Gamma1}
\end{equation}
where $g$ is some smooth differentiable function. 
This Hamiltonian depends only on the coordinates and momenta
of the extended phasespace; thus any trajectory in the extended 
phase space which satisfies Hamiltonian equations
will conserve $\Gamma$. Therefore, if a particle begins its 
motion on the hypersurface $p_0=-H(q,p,q_0)$, it will
always stay on this hypersurface.  Moreover, the trajectory of 
this particle $[q(\tau),p(\tau),q_0(\tau)]$ will
trace the values $[q,p,t]$ corresponding to the trajectory of 
the original Hamiltonian system; here $\tau$ is the
time variable which marks evolution in the extended phase space. 
The two time variables are related by
\begin{equation}
dt=g(q,q_0,p,-H)d\tau.
\label{ttau}
\end{equation}
The idea then is to integrate the system numerically in the extended phase space using the constant
timestep $d\tau$, but choosing $g$ so that the real timestep $dt$ is small during a close encounter.

To integrate the motion in a symplectic way, we need the Hamiltonian $\Gamma$ to be separable
into two parts,
\begin{equation}
\Gamma=\Gamma_A+\Gamma_B,
\label{Gamma}
\end{equation}
so that the motion due to each of $\Gamma_A$ and $\Gamma_B$ would be integrable analytically.
Once such separation is found, the system can then be integrated via a generalized leapfrog step:
\begin{equation}
U(d\tau)=U_A(d\tau/2)U_B(d\tau)U_A(d\tau/2),
\label{U}
\end{equation}
where $U_A(d\tau)$ and $U_B(d\tau)$ represent the forward motion in the extended phasespace by the
time interval $d\tau$ under the  
action of $\Gamma_A$ and $\Gamma_B$, respectively.
Preto and Tremaine 1999 and Mikkola and Tanikawa 1999 have independently found the form
of the generalized Hamiltonian which is the sum of two analytically integrable parts:
\begin{equation}
\Gamma=f(p_0+K)-f(-U[q,q_0]),
\label{ptr}
\end{equation}
where $K$ and $U$ are the kinetic and potential energies, respectively, and $f$ is an arbitrary
smooth function. The Hamiltonian in Eq.~(\ref{ptr}) is obviously of the form in Eq.~(\ref{Gamma1}),
with $g=\Gamma/(p_0+K+U)$. When $p_0=-H$, i.e.~at 
the hypersurface representing the original Hamiltonian system, $g=f^{\prime}(-U)$. The choice
$f(x)=\log x$ possesses the following surprising and useful property: the generalized leapfrog step
for a two-body problem follows {\it exactly} the correct conic-section orbit, albeit with some
time error. The real time and the time in the extended phase space are then related by
\begin{equation}
dt=d\tau/(-U).
\label{timerel}
\end{equation}
We can now see the advantage of using the extended phase space. In the extended phase space, we
can integrate the motion by using the generalized leapfrog in Eq.~(\ref{U}) with the  timestep fixed,
which is required for energy conservation.  However, the timestep in real phase space will
not be constant: when the potential energy becomes large during a close encounter, the timestep in
real space becomes small and the precision of real-space integration increases.

For our problem however, the generalized Hamiltonian in Eq.~(\ref{ptr}) is not optimal.
We are interested in a restricted 3-body problem where a star is interacting
with two much heavier Black Holes. The mass ratio between SgrA* and the IMBH is 
$1000$ or more\footnote{The mass of the IMBH is limited from above by the mass of massive stars
participating in the  core collapse of the IMBH's parent cluster, about $0.1$\% of the cluster's mass.}
and our star is initially bound to IMBH. The potential energy of the star per its unit mass is given by
\begin{equation}
U=-{M\over\left|\vec{r}-\vec{r}_1\right|}-{m\over\left|\vec{r}-\vec{r}_2\right|},
\label{u2}
\end{equation}
where $\vec{r}$ is the position vector of the star, $M$ and $\vec{r}_1$ are the mass and
the position vector of SgrA*, and $m$ and $\vec{r}_2$ are those of the IMBH. In our simulations, the star is 
treated as a massless
particle and the orbits of the two black holes are introduced by hand, with the inspiral prescription motivated
by analytical calculations. We see from Eq.~(\ref{u2}) that the potential energy of the star is dominated by
its interaction with SgrA*, even when it is well inside the Roche Lobe of the IMBH. Thus if 
the Hamiltonian of Eq.~(\ref{ptr}) is used to generate the integrator, the following paradoxical situation
may occur: the star may be on a bound orbit around the IMBH or it may be experiencing  a close encounter with the IMBH,
yet the choice of timestep will be dominated by the star's interaction with SgrA* and not with the IMBH.
Clearly, to make the integrator efficient, we need to reduce the relative contribution to the timestep from the
star--SgrA*  interaction. The following generalized Hamiltonian, found by Mikkola and Wiegert in 2002, does
the trick:
\begin{equation}
\Gamma=\log \left(p_0+K-{1-\alpha\over \left|\vec{r}\right|}\right)-
        \log \left({m\over\left|\vec{r}-\vec{r}_2\right|}+{\alpha\over \left|\vec{r}\right|}+\delta\right).
\label{mw}
\end{equation}
 Here all the position vectors are measured relative to the barycenter of black-hole binary, which
is very close to SgrA*, $\alpha\ll 1$ is a positive number, and the term $\delta$ is given by
\begin{equation}
\delta=M\times\left({1\over \left|\vec{r}-\vec{r}_1\right|}-{1\over \left|\vec{r}\right|}\right).
\label{delta}
\end{equation}
 Since we treat  the star as a test particle
and introduce the orbits of both black holes  by hand, the vectors $\vec{r}_1$ and $\vec{r}_2$ are time-dependent
and therefore are prescribed functions of $q_0$.
Both components of the Hamiltonian in Eq.~(\ref{mw}) are easily integrable: the first term on the RHS
describes Keplerian motion around the barycenter and the second term describes the motion where all
positions stay constant and all momenta change linearly with time; therefore we can efficiently construct
the generalized leapfrog operator for this Hamiltonian.
The timestep in real phase space is then given by
\begin{equation}
dt=d\tau\times\left({m\over\left|\vec{r}-\vec{r}_2\right|}+{\alpha\over \left|\vec{r}\right|}+\delta\right)^{-1}
\label{dt1}
\end{equation}
If one chooses  $\alpha=(m/M)^{2/3}$, then at the IMBH Roche surface the star-SgrA* interaction and 
star-IMBH interaction contribute approximately equally to the timestep, and the contribution from the $\delta$-term
is relatively small. 

Mikkola and Wiegert (2002) have proposed  to use the generalized Hamiltonian in
Eq.~(\ref{mw}) for simulating the motion of near-Earth and near-Jupiter asteroids. They have cautioned
however, that if the asteroid gets too close to the sun, the $\delta$-term in Eq.~(\ref{mw}) may become large,
and the Hamiltonian may become singular. Our simulations have confirmed this. We have found that while
for circular inspiral the Hamiltonian in Eq.~(\ref{mw}) always works well, for an eccentric inspiral some
of the simulation runs crash because the Hamiltonian becomes singular. However, we have identified a way to fix this
singularity problem by a slight modification of both parts of the generalized Hamiltonian:
\begin{eqnarray}
\Gamma&=&\log \left(p_0+K-{1-\alpha\over \left|\vec{r}\right|}+{\epsilon^2\over 2r^2}\right)-\nonumber\\
      & &\log \left({m\over\left|\vec{r}-\vec{r}_2\right|}+{\alpha\over \left|\vec{r}\right|}+\delta
        +{\epsilon^2\over 2r^2} \right).
\label{yl}
\end{eqnarray}
The above Hamiltonian is  of the form in Eq.~(\ref{Gamma1}) and, moreover, the two terms on the RHS are
still separately integrable. The latter statement is not trivial for the first term on the RHS; however,
recall that for a particle motion in a spherically symmetric potential $V(r)$ the effective potential
for the radial motion is given by   $V(r)+(1/2)L^2/r^2$, where $L$ is the particle's angular momentum.
Thus adding the term $(1/2)\epsilon^2/r^2$ to the Hamiltonian is equivalent to  rescaling of the particle's angular momentum
in the radial equation of motion: $L\rightarrow\sqrt{L^2+\epsilon^2}$, and  solving the equations of motion
with this term can be reduced to solving a purely Keplerian problem\footnote{We will discuss elsewhere how we have
implemented this procedure in practice.}. It is straightforward to show that by setting
the value of $\epsilon$ so that $\epsilon^2/2>mr_2^{\rm max}$ one can avoid the singularity in the
generalized Hamiltonian; here $r_2^max$ is the maximal distance from the IMBH 
to the barycenter during the simulation run.

In this work, we have used the Hamiltonians in Eq.~(\ref{mw}) and Eq.~(\ref{yl}) to simulate 
the circular and eccentric IMBH inspirals, respectively. 
 We have also, as a check,  have run test simulations
with  SYMBA, a version of widely available SWIFT code  with the added ability
to
 resolve close encounters between massive objects (Duncan, Levison, and Lee 1998). 
We have confirmed that our integrator and SYMBA give results which
are consistent with each other. 
The efficiency of our code has allowed us to run thousands of inspiral simulations
and obtain a distribution of the orbital parameters of the stars at the end of the inspiral.
In the next section, we discuss the results of our simulations.

\section{Results}
\subsection{Parameters of the inspiral.}
We show in the Appendix that the eccentricity of the
inspiraling IMBH remains nearly constant for the observed density
profile of the stellar cluster. The semimajor axis evolves with time as
\begin{equation}
a(t)=a_0\exp{(-t/t_0)},
\label{a1}
\end{equation}
where $t_0$ is the characteristic inspiral time given by
\begin{equation}
t_0={M^{1.5}\over 8\pi\sqrt{G} m\rho r^{1.5} \log(\Lambda)},
\label{t0}
\end{equation}
where $M$ is the mass of SgrA*, $m$ is the mass of the IMBH, 
$\rho$ is the mass density of stars in the central cluster,
and $\ln(\Lambda)$ is the Coulomb logarithm. For circular orbit,
this can be rewritten as 
\begin{equation}
t_0={T_{\rm orb}\over 16 \pi^2 \log(\Lambda)}\times\left({M\over m}\right)
    \left({M\over\rho r^3}\right),
\label{t01}
\end{equation}
where $T_{\rm orb}$ is the orbital period of the IMBH. Genzel et.~al.~(2003)
have measured the density profile of the central cusp to be
\begin{equation}
\rho\simeq 10^6 (r/10'')^{-1.3} M_{\odot} \hbox{pc}^{-3}.
\end{equation}
For a realistic IMBH  $M/m\sim 1000$.
Therefore for the stars located a few arcseconds from SgrA* the
relevant inspiral timescale is between a few 
hundred and a few thousand orbital periods. In our simulations
we use fiducial values of the  mass ratio $M/m=1000$ and the inspiral timescale
of one thousand initial orbital periods of the IMBH. Our inspiral timescale remains
constant over the simulation run; this approximation  would be exact if the cusp density profile
scaled as $\rho\propto r^{-1.5}$ (see e.g.~Appendix of Gould and Quillen, 2003). We  spot-check our results
by varying the inspiral timescale and making sure that our qualitative conclusions do not change.

\subsection{Circular inspiral.}

Figure 1 shows a pictorial representation of a typical circular inspiral.
The points $(x,y)$ on the figure represent ten thousand
 snapshots of the $x$ and $y$ coordinate
of the star. The orthogonal rotating axis $x$ and $y$ belong to the inspiral plane and are
 chosen so that the SgraA* and IMBH are always on the $x$ axis. 
From Figure 1 we can trace the dynamical history of the star as the inspiral proceeds.
Initially the star is bound to the IMBH which is climbing towards the center on
the $x$-axis. Then the tidal field of SgrA*  disrupts the star-IMBH binary, the star
escapes from the IMBH via the inner Lagrange point and spends some time inside of the
IMBH orbit. Then as a result of a close encounter, the star is flung outside of the IMBH orbit.
The star then continues to have encounters with the IMBH, and its eccentricity and inclination 
get kicks during the encounters. Eventually, the orbit of the IMBH is shrunk enough
so that the encounters do not occur any more and the 
star's eccentricity and inclination stay virtually constant.
Figure 2 and 3 show how the star's eccentricity
and inclination evolve with time.

Using CITA's Mackenzie cluster, we have performed one thousand simulation runs, such as the
one shown in Figures 1,2, and 3. The initial conditions were different for each run; 
we have chosen the stars in all the runs to have the same initial Jacoby integral.
In Figure 4 we show in the $(e,i)$ plane the scatter plot of the final eccentricities and
inclinations.  
Because of multiple close encounters, the orbits are highly chaotic: even a slight
change in the initial conditions or in the timestep of the integrator leads, after some time,
to a very different trajectory and thus to a different final eccentricity and inclination
relative to the inspiral plane. One subtle issue is the choice of timestep\footnote{Here
we mean the timestep $d\tau$ in the extended phase space; see the previous section.}.
Usually, one can make sure that the chosen timestep is appropriate by running the simulation
with the fraction of the chosen timestep and checking if the simulation results stays the same.
In our case however, the system is chaotic and the individual trajectory will always be different
after some time if a different integration timestep is chosen. However, it is not the individual
trajectories that we are after; rather, we want to know what {\it distribution} of the 
orbital parameters one should expect in the inspiral scenario. Therefore, we have repeated the
$1000$ runs with the same initial conditions
 but with the timestep of $0.3$ of the original timestep, and checked if the two sets of $(e,i)$
points could belong to the same two-dimensional distribution. The first thing
to check was whether the averages agreed; they did. However, we wanted
a more complete test. For one-dimensional data,
Kolmogorov-Smirnov test is both popular and mathematically well justified. For two-dimensional
data, there exist only semi-empirical algorithms based on Monte-Carlo simulations: see
Peacock (1983) and Fasano and Franceschini (1987). We  used the procedure outlined in the latter to check
whether the  two  $(e,i)$ data sets were compatible with the same distribution; they were.

The remarkable thing about the scatter plot in Fig.~4 is that while the stars,
on average, end up with significant eccentricity, more than 99\% of them
have inclinations smaller than 10 degrees. 
We make another
1000 runs
with the inspiral rate reduced by a factor of $10$, we find that while the
average eccentricity has increased significantly, and  the distribution
of inclinations remains similar, with only a few greater than 10 degrees.

\subsection{Eccentric inspiral.}
The final inclinations are no longer small when the inspiraling IMBH is
on an eccentric orbit. Qualitatively, eccentric stellar orbits experience
secular torque if the semimajor axes of the star and IMBH are misaligned.
Hence angular momentum,  inclination, and eccentricity  of the star can experience
substantial secular changes during the course of the inspiral. In Figure 5 we
plot a fraction of stars with inclinations greater than 10 degrees for a few values 
of IMBH eccentricity. Each point of the graph is obtained by simulating 100 stellar
orbits, and the stars which become unbound from SgrA* are removed from the data set.
 In Figure 6, we show average inclination of stellar orbits after the inspiral,
for the same values of IMBH eccentricity. In Figure 7, we show the scatter plot of final inclinations
and eccentricities of stars for the  case when the IMBH eccentricity is $0.6$; the scatter plot
contains $1000$ points. We observe  two groups of 
unejected stars which are roughly equal in size. In one group, the inclinations are less than 10 degrees,
and in another the inclinations are spread  evenly between 10 and 180 degrees.
In Figures 8 and  9 we show the time evolution 
of inclination and  eccentricity for a typical star which ends up with high
inclination.

\section{Discussion and comparison with the data.}

Current kinematic data indicates that there are two nearly
orthogonal
stellar discs in the Galactic Center. One of the disks
consists of clockwise-moving stars at $\sim 0.1pc$ from
the Black Hole, and has dispersion in
inclinations no larger than 10 degrees (Levin and Beloborodov 2003,
Genzel et.~al.~2003). Some of the stars in this clockwise
disc have significant eccentricities (Beloborodov and Levin, 2005).
The second disc consists of stars which
move counter-clockwise (Genzel et.~al.~2003) and its thickness
has not been estimated yet; however it is probably thicker than
the clockwise disc since the planar fit to its stellar velocity
vectors has a large $\chi^2$ of $3.5$. The mini-cluster IRS13 with
a putative IMBH belongs to this second, thicker disc.

In our simulations the circular inspiral nicely reproduces the kinematics
of the thin clockwise disc: dispersion of inclinations is less than 10 degrees yet   
some of the stars possess significant eccentricities. 
However, the nearly circular inspiral would never produce the second population
of stars which move counter-clockwise. Eccentric inspiral seems to be more
promising: while significant fraction of stars remain close to the inspiral plain,
the rest of the stars get pushed out of the inspiral plain and end up with
large inclinations; see Figures 5 and 7. However, the stars with large inclinations do not
have a preferred orbital orientation, thus it is hard to imagine them assembling a second
disc. Upcoming observations will better constrain the thickness of the second disc and then
it will become possible to make a more quantitative assessment of
how likely it is to produce the observed counter-clockwise trajectories by
an IMBH inspiraling through the plane of the clockwise disc.
It seems certain, though, that an IMBH in a counter-clockwise moving IRS13 would
not be able to produce the thin clockwise disc. Thus
the presence of IMBH in IRS13 could not account for kinematics for all of the young
stars at $\sim 0.1$pc from SgrA*.

The IMBH in IRS13 would also have difficulty generating highly compact eccentric orbits of
the most central stars like SO-2 or SO-16 (see Ghez et.~al.~2005). These stars' apoapses are
an order of magnitude closer to SgrA* than IRS13. Such disparity
of stellar and IMBH orbits is never observed in our simulations.
However, we note from Figure 9  that
 in the case of an eccentric inspiral, the test bodies
with high inclination undergo secular evolution in which their eccentricity reaches
values close to 1, and their distance of closest approach to the central Hole is much smaller
than their semimajor axis. Thus, if such a test body were a binary, this binary could
get disrupted by SgrA*, and one of the components of the binary could remain tightly bound
to SgrA*. The binary-disruption scenario for formation of the SO-stars was already proposed by
Gould and Quillen (2003) in a different context.

\acknowledgments
We thank Scott Tremaine for introducing us to symplectic integrators in
the extended phase space, and Carlo Contaldi for showing us how to use the
CITA McKenzie cluster. We have benefited from discussions with Andrei Beloborodov,
Jeremy Goodman,
Brad Hansen,
Chris Matzner, Milos Milosavljevic, and especially  with Norm Murray. 
Our research was supported by NSERC.

\section*{Appendix:  IMBH eccentricity evolution during the inspiral.}

The orbit of an IMBH in the Galactic Center shrinks due
to dynamical friction, and its eccentricity may also evolve with
time. The details of this evolution depend on the structure of the stellar
cusp surrounding the Black Hole. It is instructive and convenient to model
the cluster by a collection of  stars of mass $m_*$ with an isotropic distribution function
\begin{equation}
f(r,v)=\beta E^{\gamma},
\label{isot}
\end{equation}
where $E=GM/r-v^2/2$ is the binding energy of the star, and $\beta$ and $\gamma$ are real numbers.
For a Keplerian orbit, the angular momentum and binding energy are given by $L=\sqrt{GMa(1-e^2)}$
and $E=0.5GM/a$, respectively; here $e$ is the eccentricity of the orbit. From these relations
it is straightforward to derive the evolution equation for IMBH eccentricity:
\begin{equation}
{d\log{e}\over dt}= -{1-e^2\over 2e^2}\left\langle
{d\log E\over dt}+2{d\log L\over dt}\right\rangle.
\label{eevol}
\end{equation}
Here the symbol $\langle\rangle$ represents averaging over the orbital period. The 
acceleration from dynamical friction  acting on the IMBH with velocity
vector $\vec{v}$ is given by the Chandrasekhar formula
\begin{equation}
\vec{a}=-k\vec{v},
\end{equation}
where
\begin{equation}
k={4\pi G^2 m_* m\log\Lambda\over v^3}\int_0^v du\times 4\pi u^2
  f(r,u).
\label{kc}
\end{equation}
Therefore,
\begin{equation}
\left\langle dE/dt\right\rangle=\langle kv^2\rangle,
\label{dEdt}
\end{equation}
and
\begin{equation}
\langle dL/dt\rangle=-\langle k\rangle L;
\label{dLdt}
\end{equation}
cf.~Eqs (A5) and (A6) of Gould and Quillen (2003). It is convenient
to perform orbital averaging by expressing all the quantities through the mean anomaly $\phi$:
\begin{eqnarray}
r&=&a(1-e\cos\phi),\nonumber\\
v^2&=&{GM\over a}{1+e\cos\phi\over 1-e\cos\phi},\label{manomaly}\\
dt/T_{\rm orb}&=&(1-e\cos\phi){d\phi\over 2\pi}.\nonumber
\end{eqnarray}
After some algebra, one gets
\begin{equation}
{d\log e\over d\log a}=-{1-e^2\over 2e^2}(k_1/k_2-1),
\label{deda}
\end{equation}
where 
\begin{equation}
k_1=\int_0^1 x^2 dx\int_0^{2\pi}d\phi [1-(1+e\cos\phi)x^2]^{\gamma}(1-e\cos\phi)^{1-\gamma},
\label{k1} 
\end{equation}
and 
\begin{equation}
k_2=\int_0^1 x^2 dx\int_0^{2\pi}d\phi [1-(1+e\cos\phi)x^2]^{\gamma}(1-e\cos\phi)^{-\gamma}
      (1+e\cos\phi).
\label{k2} 
\end{equation}
We evaluate the integrals numerically and in Fig.~10 we 
plot $-d\log e/d \log a$ as a function of eccentricity, for 
$\gamma=-0.2$ and $\gamma=0.25$. The
two values of $\gamma$ correspond to the observed density profile
$\rho\propto r^{-1.3}$ and the Bahcall-Wolf density profile
$\rho\propto r^{-1.75}$. We see from the graphs that in both
cases eccentricity changes by a factor less than 2 as the semimajor axis
shrinks by 3 orders of magnitude. Hence, in our simulations we are justified in keeping
the eccentricity of IMBH orbit  constant.

\newpage
\begin{figure}
\begin{center}
\plotone{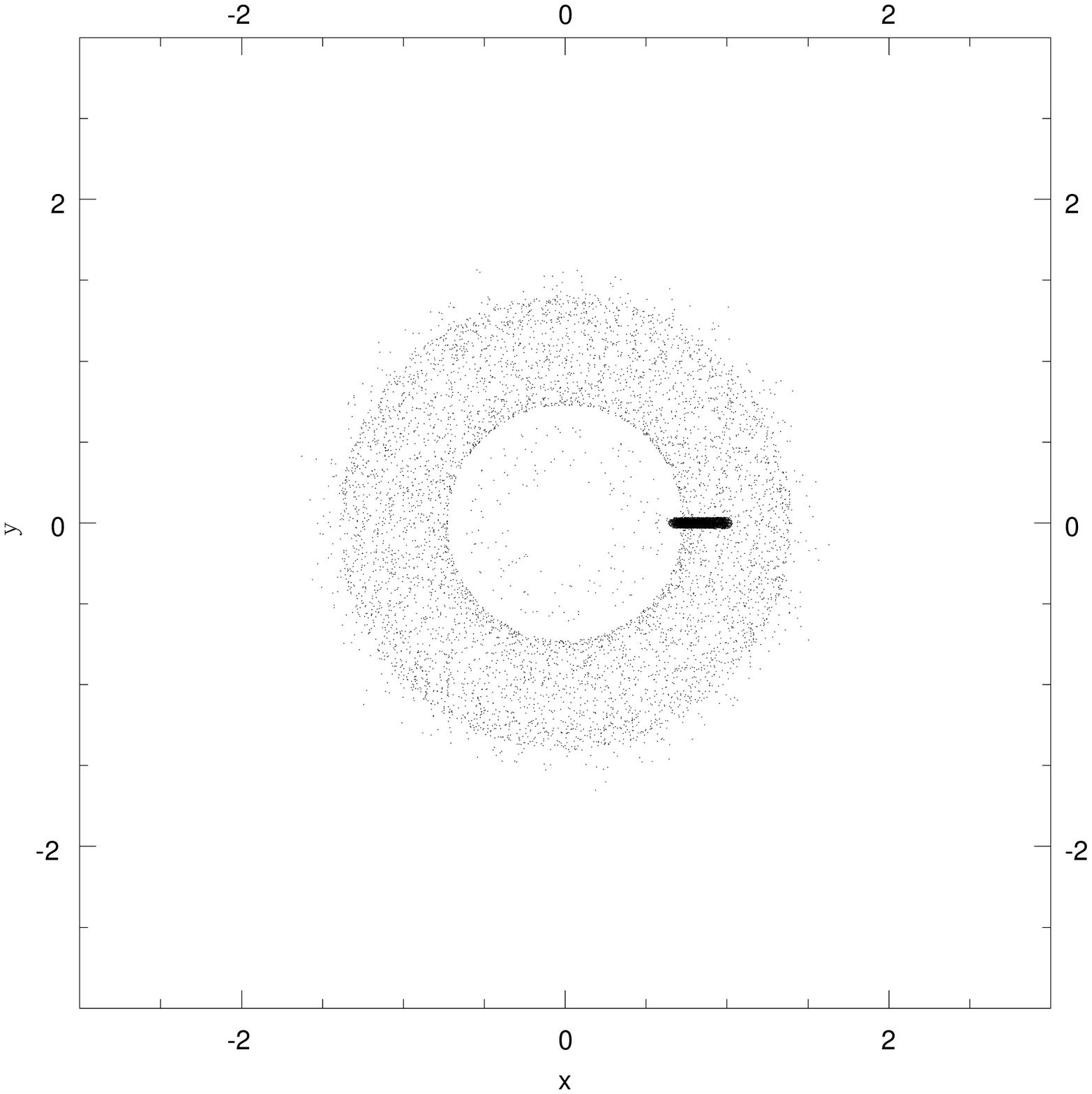}
\end{center}
\caption{10000 snapshot of the steallar position in the IMBH inspiral plane. Rotating coordinates
axes are chosen so that both IMBH and SgrA* are on the $x$-axis, and their origin coinsides with
the baricenter. The star is originally bound to the IMBH; this corrensponds to the dense dark region
of the scatter plot. The star escapes through the inner Lagrange point, then gets flung outside
by a close encounter with the IMBH. A few close encounters follow, and then the
orbit stabilises once the black-hole binary is shrunk away from the stellar periapse. Units
on $x$ and $y$ axes are arbitrary.}

\end{figure}

\begin{figure}
\begin{center}
\plotone{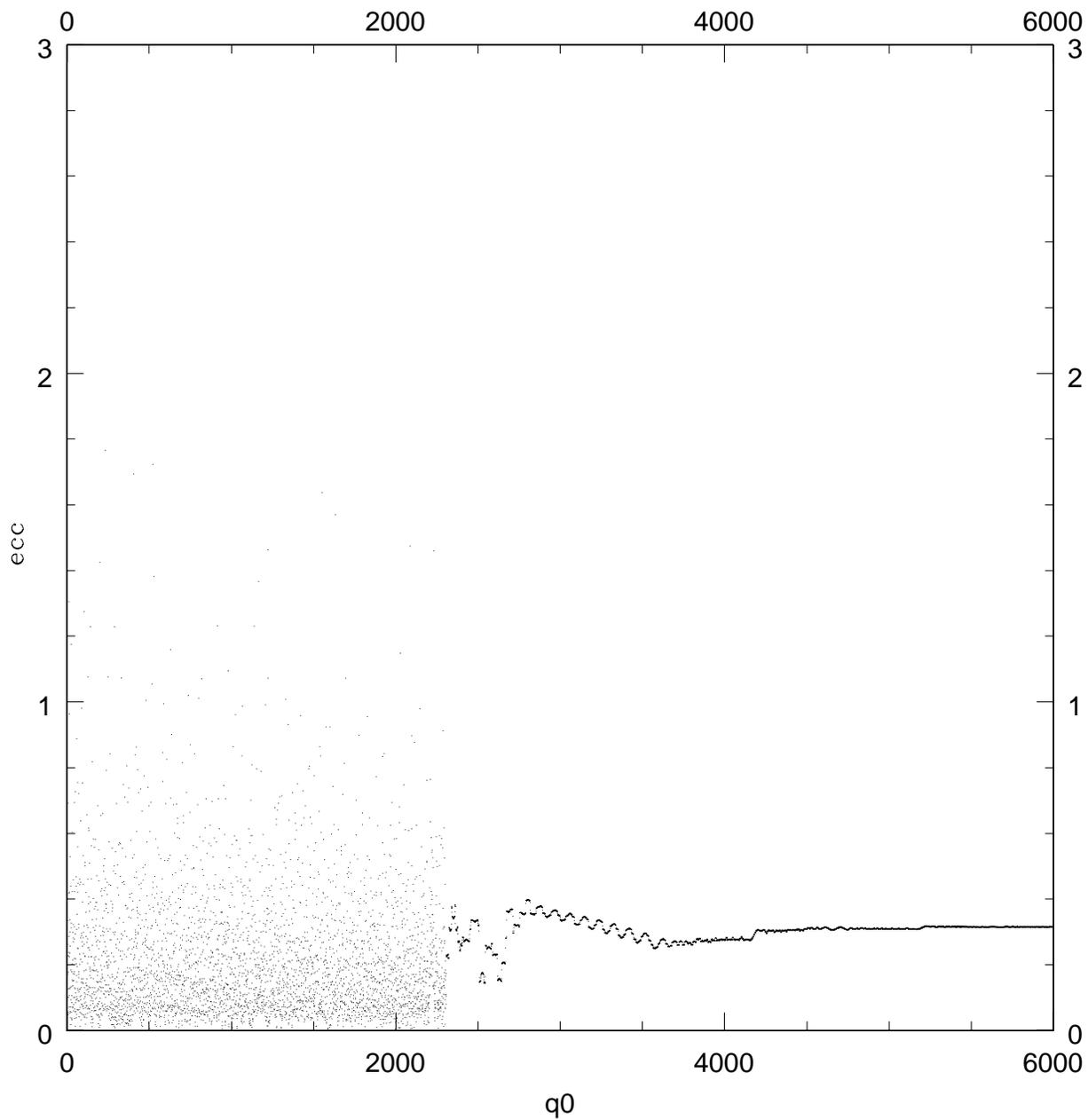}
\end{center}
\caption{10000 snapshots of eccentricity evolution for the star in Figure 1. The eccentricity is evaluated relative
to the baricenter. When the star is bound to the IMBH and comes close to it, the eccentricity
relative to the baricenter may become greater than 1. Time units: initial IMBH orbital period is $2\pi$.}

\end{figure}

\begin{figure}
\begin{center}
\plotone{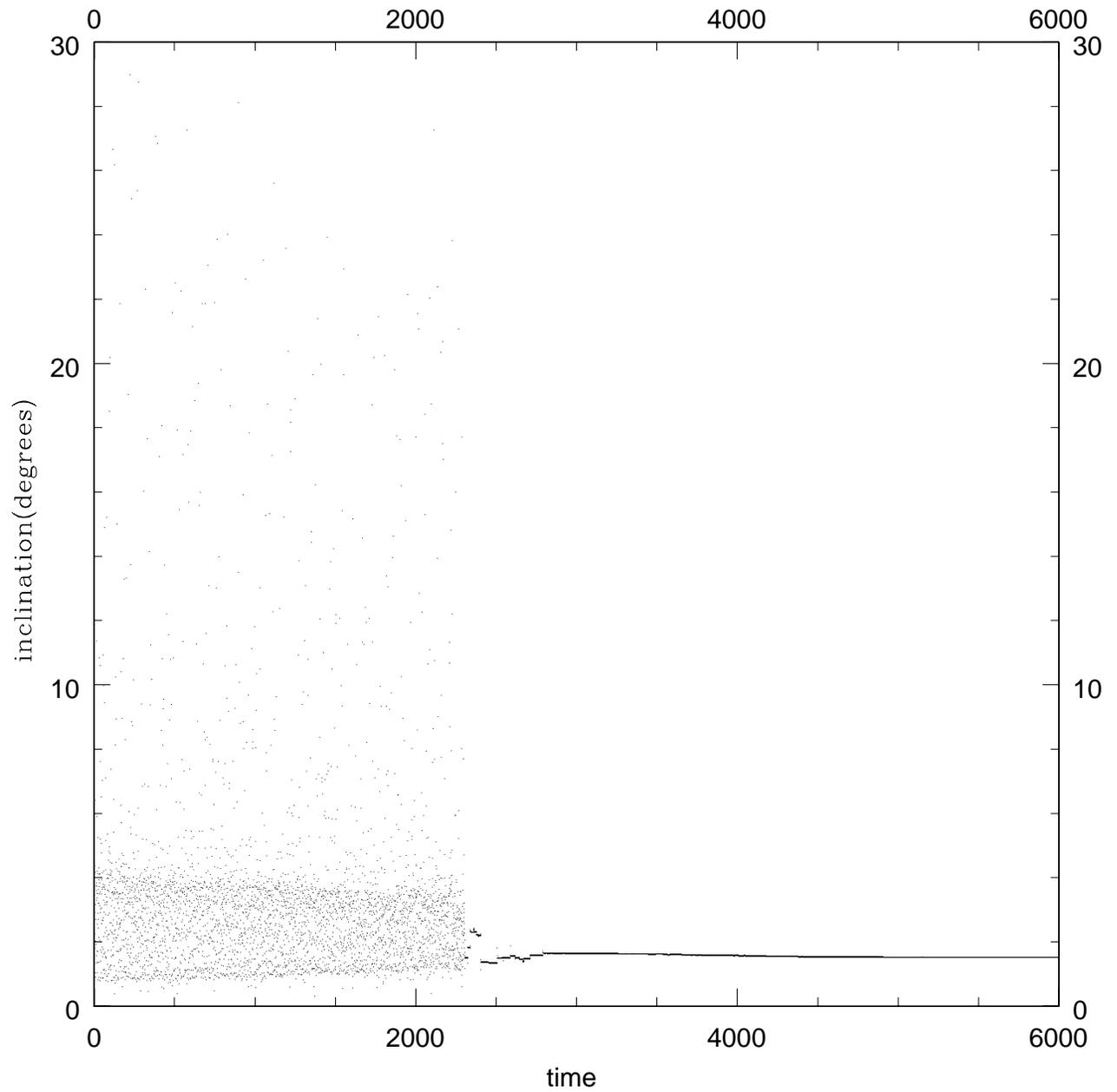}
\end{center}
\caption{10000 snapshots of inclination evolution for the star in Figure 1. The inclination is evaluated relative
to the baricenter. When the star is bound to the IMBH and comes close to it, the inclination
relative to the baricenter may become large. Time units: initial IMBH orbital period is $2\pi$.}

\end{figure}

\begin{figure}
\begin{center}
\plotone{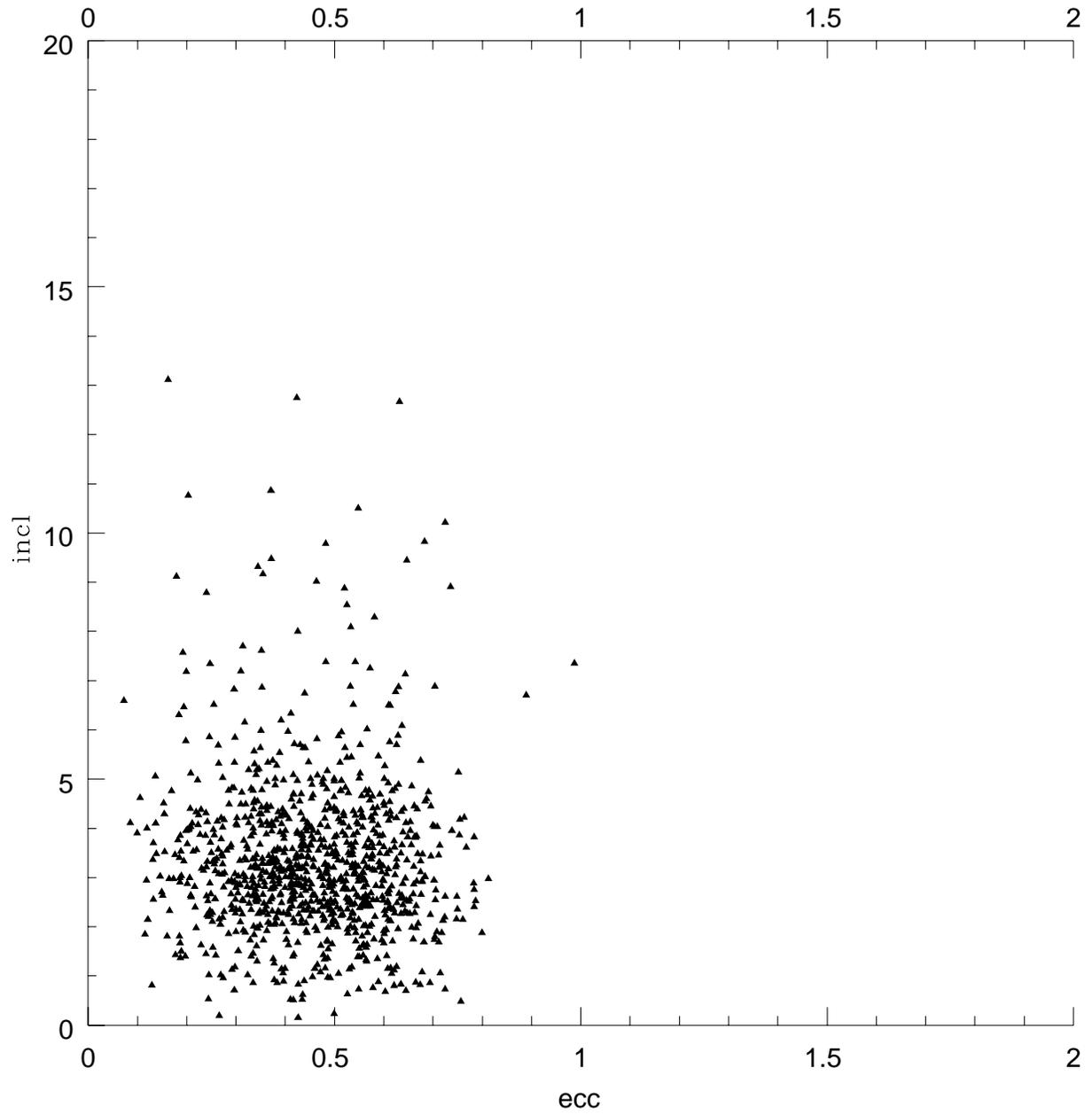}
\end{center}
\caption{Scatter plot of eccentricities and inclinations for a thousand stars after
a circular inspiral.}

\end{figure}

\begin{figure}
\begin{center}
\plotone{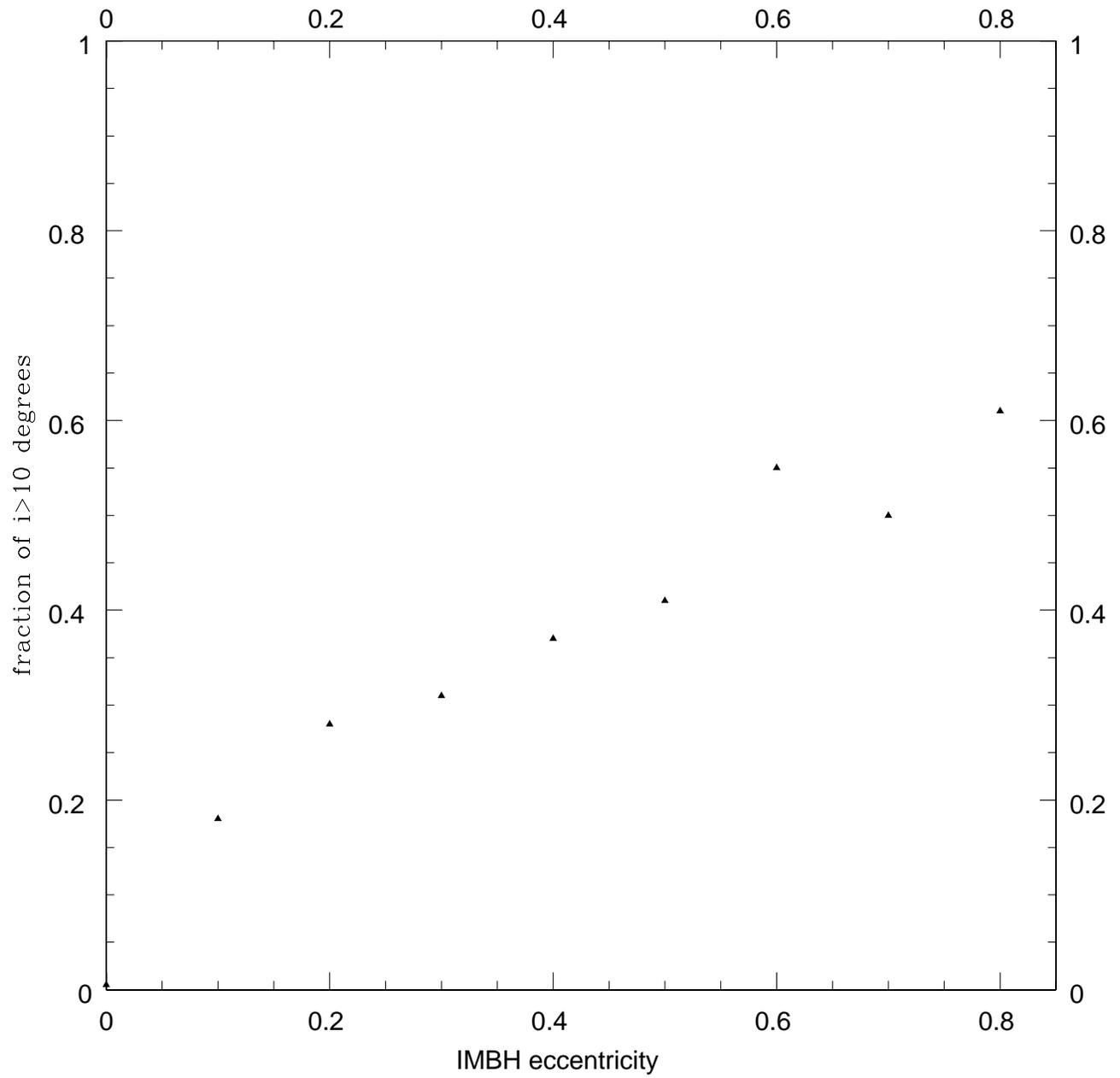}
\end{center}
\caption{A fraction of stars with inclination $>10$ degrees plotted vs IMBH eccentricity.
Each point is obtained from 100 runs.}

\end{figure}

\begin{figure}
\begin{center}
\plotone{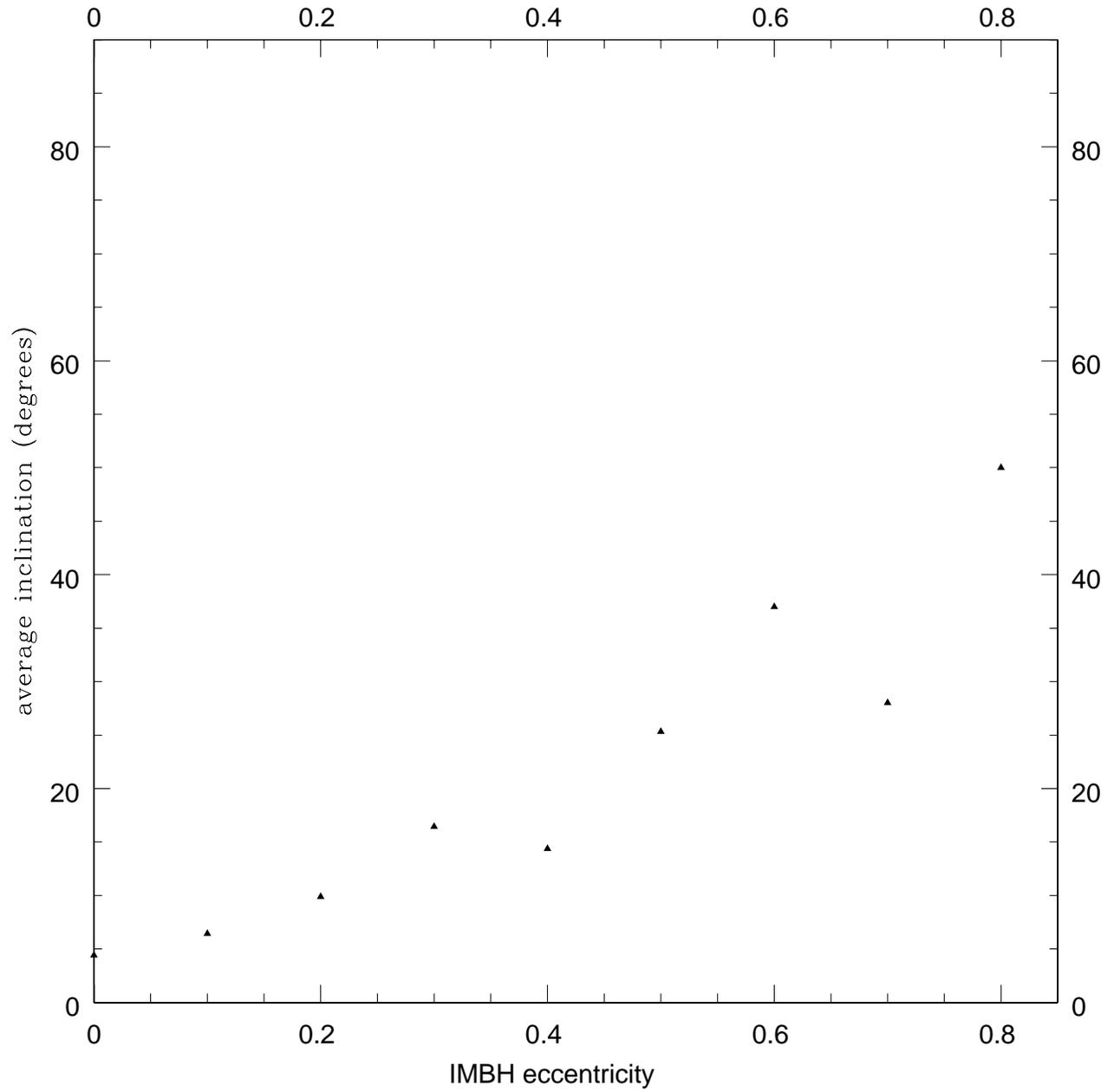}
\end{center}
\caption{Average inclination plotted vs IMBH eccentricity. Each point is obtained from 100 runs.}

\end{figure}

\begin{figure}
\begin{center}
\plotone{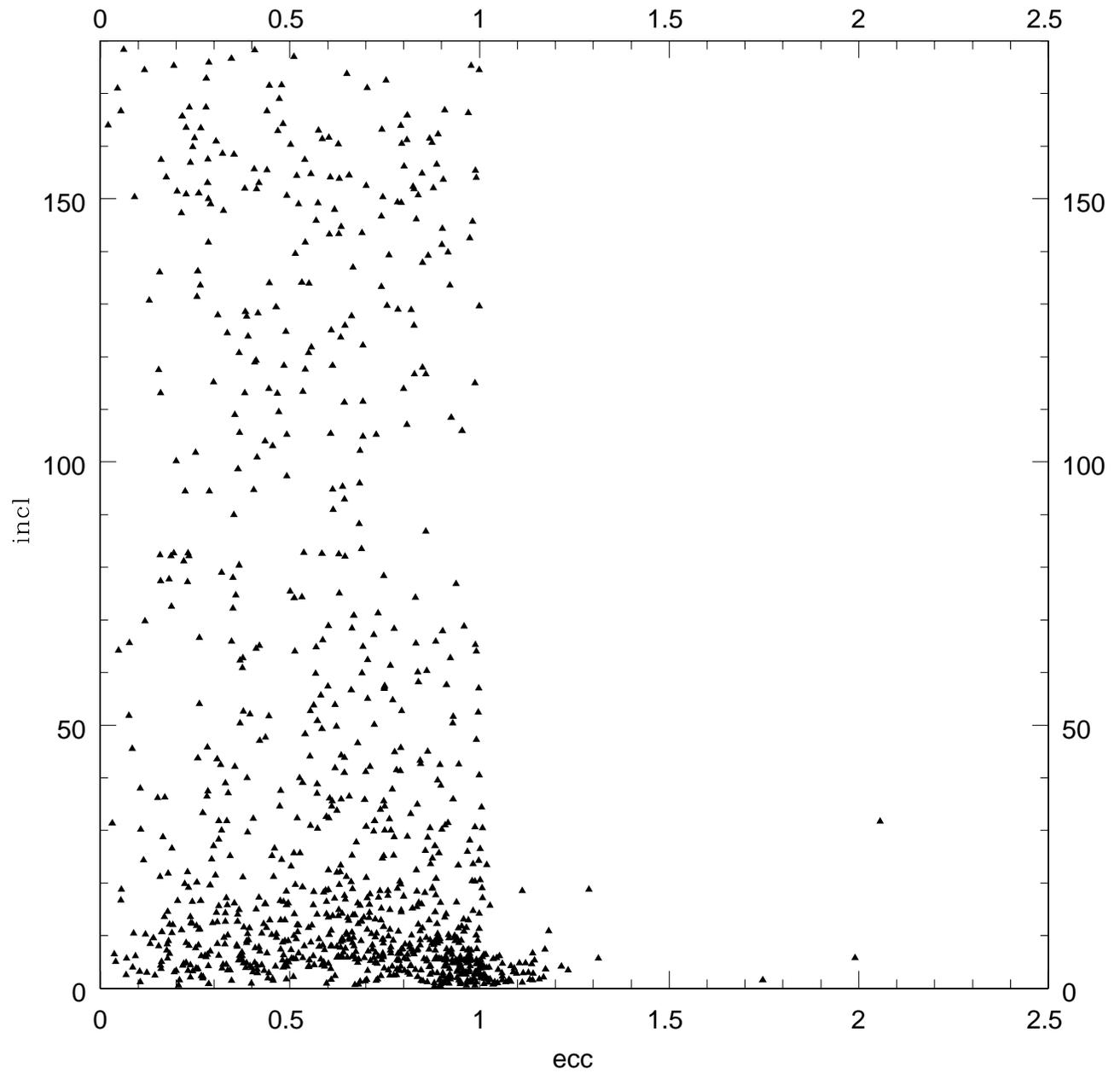}
\end{center}
\caption{Scatter plot of eccentricities and inclinations for a thousand stars after
an inspiral with IMBH eccentricity of $0.6$.}

\end{figure}

\begin{figure}
\begin{center}
\plotone{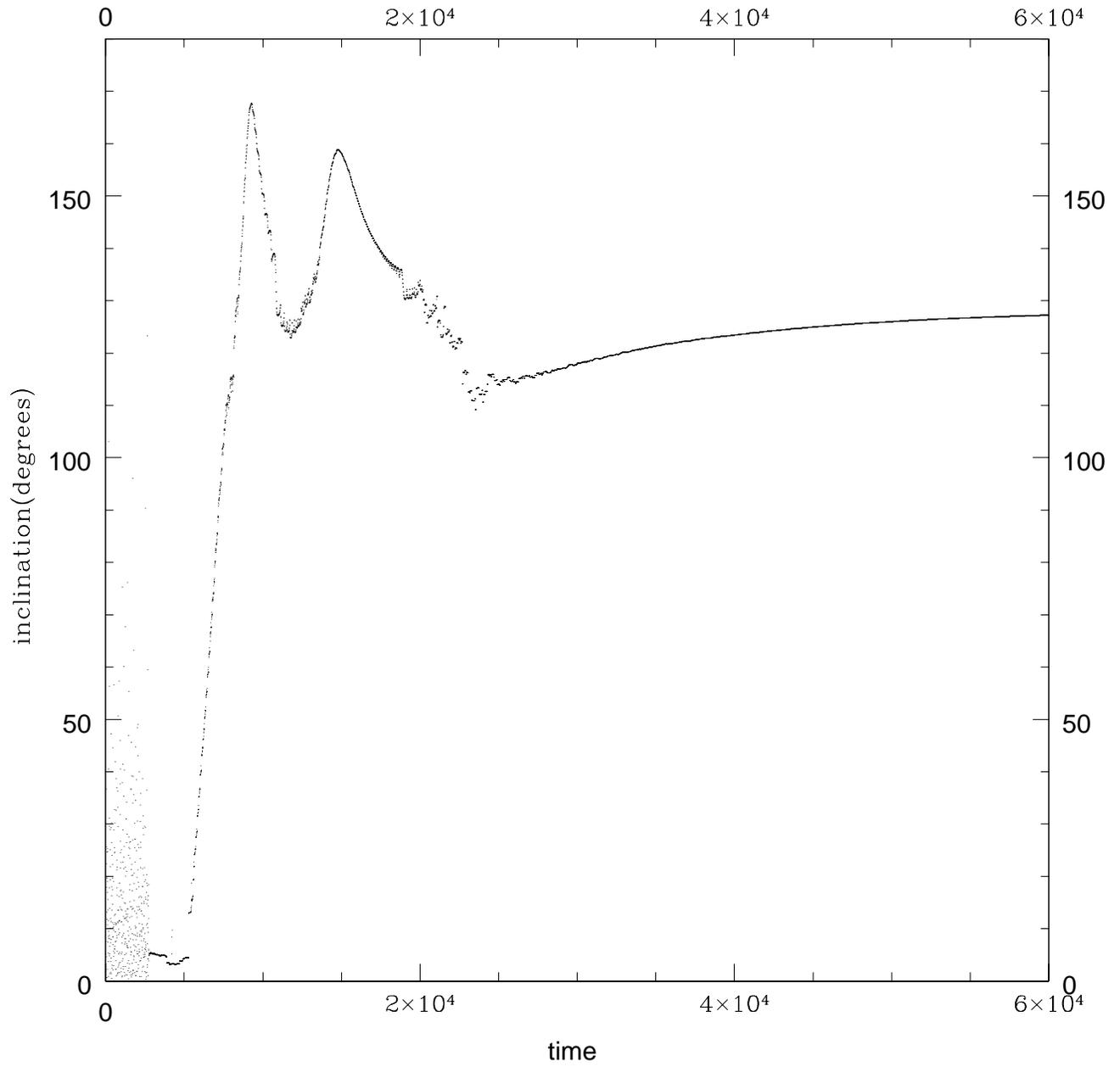}
\end{center}
\caption{Inclination evolution of a typical high-inclination star; IMBH eccentricity
is $0.6$}

\end{figure}

\begin{figure}
\begin{center}
\plotone{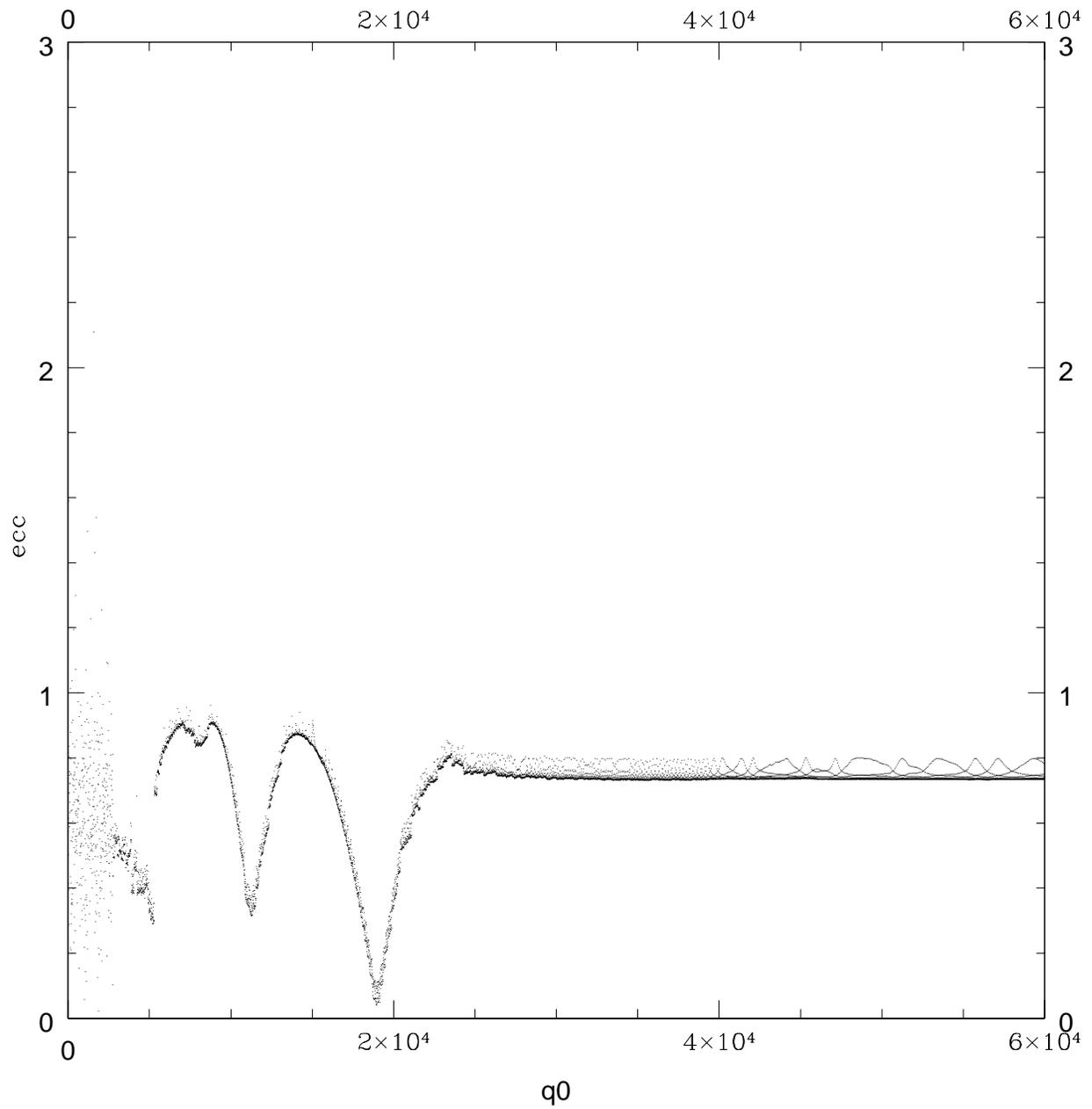}
\end{center}
\caption{Eccentricity evolution of the same high-inclination star as in Figure 8.}

\end{figure}

\begin{figure}
\begin{center}
\plotone{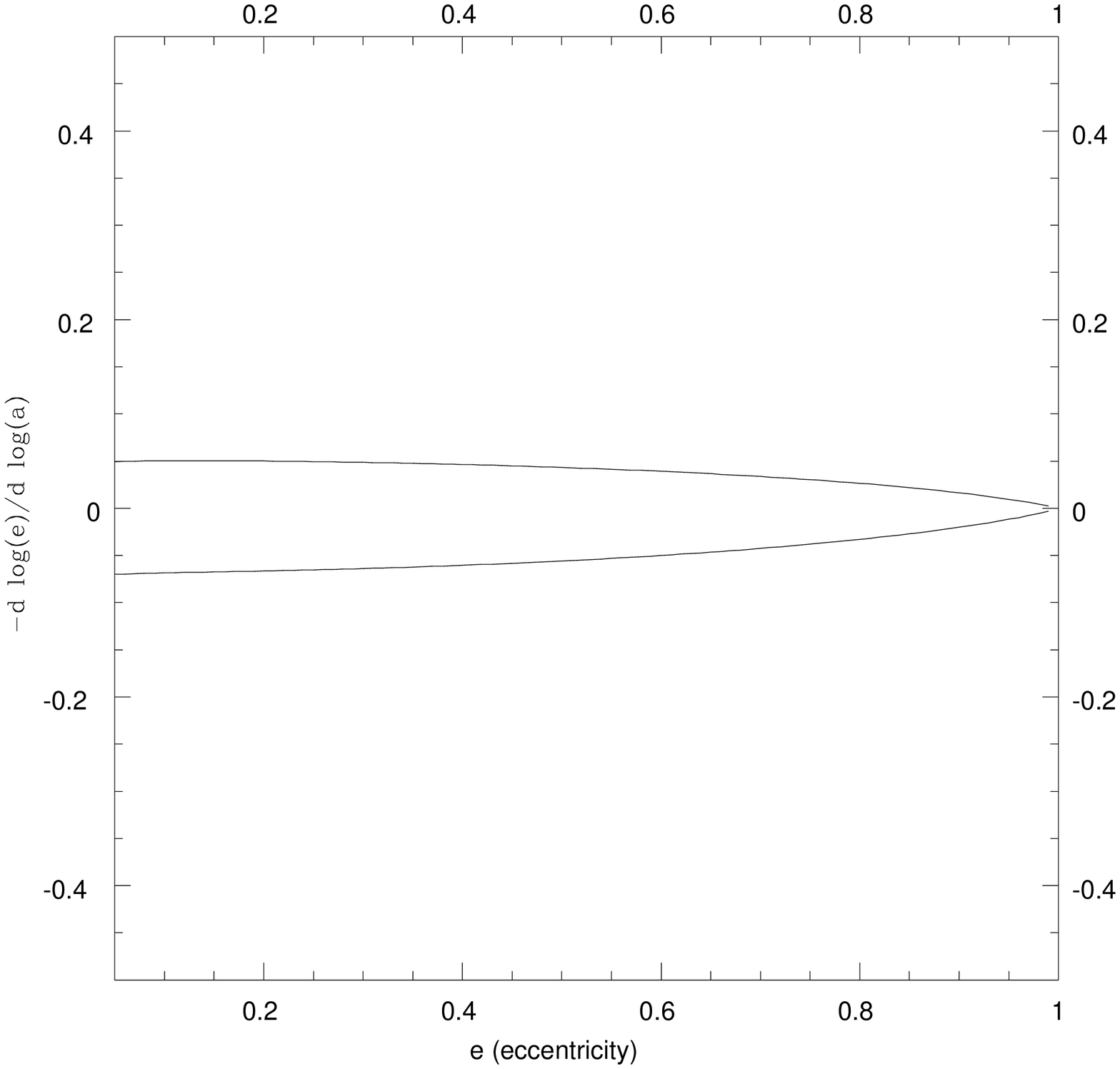}
\end{center}
\caption{ Plot of the IMBH eccentricity
growth rate $d\log e/d\log (1/a)$ vs $e$. Upper curve is computed for
the observed density profile of luminous matter in the Galactic Center, $\rho\propto r^{-1.3}$,
and lower curve is computed for a relaxed Bahcall-Wolf density profile, $\rho\propto r^{-1.75}$.}

\end{figure}


\begin{references}
\reference{} Alexander, T., \& Livio, M.~2004, 606, L21
\reference{} Duncan, M.~J., Levison, H.~F., \& Lee, M.~H.~1998, AJ, 116, 2067 
\reference{} Fasano, G., \& Franceschini, A.~1987, MNRAS, 225, 155
\reference{} Genzel, R., et.~al.~2000, MNRAS, 317, 348
\reference{} Genzel, R., et.~al.~2003, ApJ, 594, 812
\reference{} Gerhard, O.~2001, ApJ, 546, L39
\reference{} Ghez, A.~M., et.~al.~2003, ApJ, 586, L127
\reference{} Ghez, A.~M., et.~al.~2005, to appear in ApJ (astro-ph/0306130)
\reference{} Goodman, J., \& Tan, J.~C.~2004, ApJ, 608, 108
\reference{} Gould, A., \& Quillen, A.~C.~2003, ApJ, 592, 935
\reference{} Gurkan, M.~A., Freitag, M., \& Rasio, F.~A.~2004, ApJ, 604, 632
\reference{} Gurkan, M.~A.,  \& Rasio, F.~A.~2004, submitted to ApJ (astro-ph/0412452)
\reference{} Hansen, B.~M.~S., \& Milosavljevic, M.~2003, ApJ, 593, L77
\reference{} Jaroszynski, M.~2000, AcA, 50, 67
\reference{} Kim, S.~S., Figer, D.~F., \& Morris, M.~2004, ApJ, 607, L123
\reference{} Kolykhalov, P.~I., \& Sunyaev, R.~A.~1980, SvAL, 6, 357
\reference{} Levin, Y., \& Beloborodov, A.~M.~2003, ApJ, 590, L33
\reference{} Levin, Y.~2003, astro-ph/0307084
\reference{} Maillard, J.~P., et.~al.~2004, A\&A, 423, 155
\reference{} McMillan, S.~L.~W., \& Portegies Zwart, S.~F.~2003, ApJ, 596, 314
\reference{} Mikkola, S.~1997, CeMDA, 67, 145
\reference{} Mikkola, S., \& Tanikawa, K.~1999, CeMDA, 74, 287
\reference{} Mikkola, S., \& Wiegert, P.~2002, CeMDA, 82, 375
\reference{} Milosavljevic, M., \& Loeb, A.~2004, ApJ, 604, L45
\reference{} Morris, M, Ghez, A.~M., \& Becklin, E.~E.~1999, Advances in Space Research,
             23, 959
\reference{} Nayakshin, S., Cuadra, J., \& Sunyaev, R.~2004, A\&A, 413, 173
\reference{} Peacock, J.~A.~1983, MNRAS, 202, 615
\reference{} Phinney, E.~S.~1989, The Center of the Galaxy (136th symposium of the IAU,
             edited by M.~Morris), 543
\reference{} Portegies Zwart, S.~F., et.~al.~2004, Nature, 428, 724
\reference{} Preto, M., \& Tremaine, S.~1999, AJ, 118, 2532
\reference{} Sanders, R.~H.~1998, MNRAS, 294, 35
\reference{} Schodel, R., et.~al.~2002, Nature, 419, 694
\reference{} Schodel, R., et.~al.~2003, ApJ, 596, 1015
\reference{} Shlosman, I., \& Begelman, M.~1987, Nature, 329, 810
\reference{} Yu, Q., \& Tremaine, S.~2003, ApJ, 599, 1129



\end{references}
\end{document}